\documentclass  {article}

\usepackage{amsfonts}
\usepackage{amssymb}
\usepackage{graphicx}

\begin{document}

\title{%
 $\kappa$-Poincar\'e dispersion relations and the black hole radiation}
\author{A.~B\l{}aut\thanks{e-mail
address ablaut@ift.uni.wroc.pl}~, J.\ Kowalski--Glikman\thanks{e-mail
address jurekk@ift.uni.wroc.pl}~~\thanks{Partially supported by the   KBN grant 5PO3B05620}~, and D.~Nowak--Szczepaniak\thanks{e-mail
address dobno@ift.uni.wroc.pl}\\ Institute for Theoretical
Physics\\ University of Wroc\l{}aw\\ Pl.\ Maxa Borna 9\\
Pl--50-204 Wroc\l{}aw, Poland} \maketitle

\begin{abstract}
Following the methods developed by Corley and Jacobson, we
consider qualitatively the issue of Hawking radiation in the case
when the dispersion relation is dictated by quantum
$\kappa$-Poincar\'e algebra. This relation corresponds to field
equations that are non-local in time, and, depending on the
sign of the parameter $\kappa$, to sub- or superluminal signal
propagation. We also derive the conserved inner product, that can be
used to
 count modes, and therefore to obtain the spectrum of black hole radiation in this case.
\end{abstract}
\clearpage

\section{Introduction}
In the recent years there is a growing interest in investigations of the possible role
played by modified (or broken) Lorentz invariance in ultra-high energy phenomena.
 There are many reasons for that. First of all we do not have an access yet to  any
 experimental data concerning effects that take place at the energy
scale close to Planck scale. Given this and the fact that there
is something alarming in the ease one can, in principle, probe
the Planck scale just by Lorentz boosting, it is natural to ask a
question what would happen if one deforms the Lorentz (or
Poincar\'e) symmetry. Moreover, it seems quite likely that we may
have an access to the Planck scale physics in the near future,
and perhaps, we already see traces of Planck-scale phenomena in the form of cosmic rays
anomalies, that can be explained by making use of modification of
Lorentz symmetry (see \cite{gacpir}, \cite{gacluknow} and references therein.)

Most of the papers studying the problem of modified Lorentz symmetry addressed the question as to
 how modified dispersion relation would influence physical phenomena,
  which might be ``windows'' to the Planck scale physics. One of them is the structure
  formation in inflationary cosmology, where the fluctuations that we see now in
  the form of temperature fluctuation in the background microwave  radiation spectrum
   were initially to be of the size comparable with the Planck length \cite{mb},
   \cite{jkgcosm}, \cite{niem}, \cite{star}.

Another setting in which such analysis was performed is the black hole physics
 and the issue of Hawking radiation \cite{unruh}, \cite{brout}, \cite{corjac},
 \cite{cor}. This works has shown that the properties of Hawking radiation
  are highly insensitive to the class of deformations of dispersion relations,
  which has been considered. The only exception from this rule seems to be a case of
  a black hole with both inner and outer horizons considered in \cite{bhlaser}.
  One should be careful however in making any  judgment on the basis of a finite number
  of examples analyzed so far. The dispersion relation considered by Unruh \cite{unruh}
  was devised so as to mimic a property of fluid, making possible observation of
  ``sonic black holes''. The dispersion relation of Corley and Jacobson, on the other hand
  can be understood only as a leading-order approximation of the unknown dispersion
  relation governing the Planck scale physics, and thus, their analysis cannot be regarded
  as fully conclusive.

In this paper we will repeat the analysis of Corley and Jacobson
\cite{corjac} in the case of dispersion relation motivated by a
possible role that quantum algebras may play in fundamental
physics.

The rest of this paper is organized as follows. In section 2 we
introduce the $\kappa$-Poincar\'e dispersion relation and field equations in flat
spacetime that follow from it, together with a properly defined inner
product for complex scalar fields. Section 3 will be devoted to black
hole radiation in the standard case, and in section 4 we will
perform the qualitative analysis in the case of deformed dispersion relation.

\section{Field equations in Minkowski spacetime}
Our starting point will be the so-called
$\kappa$-Poincar\'e algebra \cite{lunoruto}, \cite{maru},
\cite{luruza}, being a quantum deformation of the standard
Poincar\'e algebra that in recent months has become an object of
intensive studies \cite{gac1}, \cite{gac2}, \cite{jkgminl},
\cite{gacnew}, as a possible candidate for the algebra of {\em
kinematical} symmetries of Planck scale physics. In the massless
case this dispersion relation takes the form
\begin{equation}\label{1}
 \left(2\kappa \sinh\left(\frac{\omega}{2\kappa}\right)\right)^2 - \,{\vec{k}}^2 e^{\omega/\kappa}=0,
\end{equation}
where $\omega$ and $\vec{k}$ are energy and momentum,
respectively, and $\kappa$ is the parameter of dimension of mass
(in the unit where $c$ and $\hbar$ are set equal to $1$), which is
to be identified with the Planck mass\footnote{This identification
is supported by the analysis of cosmic rays anomalies, see
\cite{gacpir}. }. At this point one should stress that
there is an important difference between the
case considered in this paper and the one analyzed by Unruh,
Corley, Jacobson and others. Namely the dispersion relation
(\ref{1}) leads to field equations which are non-local in time,
contrary to equations non-local in space studied before.

In the rest of the paper we will be dealing with the
two-dimensional case, and in order to make the problem treatable,
instead of using the dispersion relation (\ref{1}) our starting
point will be a slightly modified relation
\begin{equation}\label{1a}
 k^2 =e^{-\omega/\kappa}\left(2\kappa
 \sinh\left(\frac{\omega}{2\kappa}\right)\right)^2=\kappa^2
 \left(1 - e^{-\omega/\kappa}\right)^2.
\end{equation}
The major difference between these two relations is that (\ref{1}) is {\em invariant} with respect to $\kappa$-Poincar\'e transformations (see, e.g., \cite{luruza}), while (\ref{1a}) only covariant (i.e., invariant ``on-shell'', when (\ref{1a}) holds.)

In what follows we will consider two cases: $\kappa >0$ which will
be called super-luminal and $\kappa <0$ which will be called
sub-luminal. This terminology is motivated by the fact that in the
former case the (momentum dependent) speed of massless modes,
defined as ${\cal C} = \partial p_0/\partial p=\partial
\omega/\partial k$ for $\omega > 0$, is greater than $1$, and
smaller than $1$ in the latter (see \cite{gac1} --
\cite{rbgacjkg}).
\newline

In order to compute a spectrum of black hole radiation one must
have in disposal a conserved inner product, whose existence in the
case of a system with infinite number of time derivatives is by no
means clear, and moreover, even if such a product exists one must
be able  to explicitly find mutually orthogonal modes of positive and negative norms.
 Let us show therefore how such a product can be constructed in two
dimensional Minkowski spacetime.

We start with the following equation of motion for  the complex
scalar field $\phi$,
\begin{equation}\label{xa}
  f(i\partial_t)\phi +\partial_x^2\phi =0,
\end{equation}
and its complex conjugate
\begin{equation}\label{xacc}
  f(i\partial_t)^*\phi^* +\partial_x^2\phi^* =0,
\end{equation}
for some analytical function $f$, whose exact form will be given in (\ref{xc})
below. In order to construct the scalar product, we start with the
following integral for the two arbitrary complex scalar fields
$\phi_1$, $\phi_2$
$$
\int_{D}\, dxdt\left\{\phi_1^* \left(f(i\partial_t) +
\partial_x^2\right)\phi_2 - \phi_2\left( f(i\partial_t)
+\partial_x^2\right)^*\phi_1^*\right\}
$$
 where $D$ denotes a compact integration region, which is bounded
 by two space-like hypersurfaces  at $t_1$ and $t_2$, say. The
 expression under integral is a divergence $\sim
 \partial_t Q + \partial_x J$. Now if we assume that the functions
 $\phi_i$ are on-shell and have a proper fall-off at spacial infinity, one can
define the conserved inner product as
\begin{equation}\label{xb}
  \Omega(\phi_1, \phi_2) = -i\int_{t=const} \, dx\, Q(\phi_1, \phi_2).
\end{equation}
To find $Q$ let us first expand
\begin{equation}\label{xc}
 f(z) = \sum_{n=0}^\infty a_n z^n, \; \; a_n\in \mathbb{R}.
\end{equation}
Then it can be easily shown by partial integration that $Q(\phi_1, \phi_2)$ is of the form
\begin{equation}\label{xd}
  Q(\phi_1, \phi_2) = \sum_{n=1}^\infty i^n a_n
  \sum_{k=0}^{n-1} (-1)^k
  \partial_t^k\phi_1^*\partial_t^{n-k-1}\phi_2.
\end{equation}
Let us now show that plane waves solutions form an orthogonal
system with respect to the product (\ref{xb}). Let $$\phi_1 = e^{i
k_1 x - i \omega_1(k_1) t}, \quad \phi_2 = e^{i k_2 x - i
\omega_2(k_2) t}.$$ Then
\begin{equation}\label{xe}
  \Omega(\phi_1, \phi_2  ) = 2\pi \delta(k_1-k_2)\sum_{n=1}^\infty a_n
  \sum_{k=0}^{n-1} \omega^k_1\omega_2^{n-k-1}.
\end{equation}
The norm of a plane wave is therefore
\begin{equation}\label{xf}
 \Omega (\phi, \phi  ) \sim 2\pi \sum_{n=1}^\infty a_n
  \sum_{k=0}^{n-1} \omega^k\omega^{n-k-1} =  2\pi \sum_{n=1}^\infty n\, a_n\, \omega^{n-1}=
  2\pi
  \frac{d\,f(\omega)}{d\, \omega}.
\end{equation}
The waves with different $k$ are orthogonal, of course. There is
 however one more case which must be considered, namely what happens
 when one has to do with a
single $k$ but two different $\omega$. In the standard case,
$\vec{k}{}^2 = \omega^2$ we have $\omega_1 =-\omega_2$, in the
expansion (\ref{xc}) only $a_2=1$ differs from zero, and the
situation is simple. In the case at hands, the reasoning is bit
more involved.
 Since $\omega_1 \neq
-\omega_2$  are related to the same $k$, using (\ref{xc}) one can
write
$$k^2 = \sum_{n=0}^\infty a_n \omega_1^n = \sum_{n=0}^\infty a_n
\omega_2^n,$$ from which  we obtain $$0=\sum_{n=0}^\infty a_n
(\omega_1^n - \omega_2^n) = (\omega_1 - \omega_2)\sum_{n=0}^\infty
a_n \sum_{k=0}^{n-1} \omega_1^k\omega_2^{n-k-1}.$$ Thus we
found that in general
\begin{equation}
\Omega( \phi_1,\phi_2) =2\pi\delta(k_1-k_2)\delta_{\omega_1,\,
\omega_2}\frac{d\,f}{d\,\omega}(\omega_1),
 \end{equation}
and this enables us to find the pseudo-orthonormal basis for the
inner product (\ref{xb}), which will be used in the analysis of
the spectrum of the black hole radiation in the case of modified
dispersion relation (\ref{1a}).
\newline

\section{Hawking radiation, standard case}

Here we  recall briefly the step leading to derivation of Hawking
radiation in the standard case, i.e., when
$k^2=f(\omega)=\omega^2$.

Following Corley and Jacobson we consider the (two dimensional) black hole metric
of the form
\begin{equation}\label{4}
  ds^2 = dt^2 - (dx - v(x) dt)^2,
\end{equation}
where for the Schwarschild spacetime $v(x) = - \sqrt{2M/(x+2M)}$.

Consider  a single frequency WKB mode of the form
\begin{equation}\label{b}
  \phi \sim \exp\left( i \int k dx\right)\, e^{-i\omega t},
\end{equation}
where $\phi$ is a massless Klein-Gordon field in the metric
(\ref{4}).
  Assuming that both $k$ and $v$ are slowly varying with position we get  the dispersion relation of
     the form
\begin{equation}\label{a}
 (\omega - vk)^2 =k^2,
\end{equation}
 from which $k = \omega/(1+v)$. Observe that this equation defines a frequency
 $\omega' = \omega - vk$ being the frequency of the wave as seen by the freely falling observer.
 We seek the minimal (negative) value of $v$ for which eq.~(\ref{a}) has a solution.
 In the case at hands we take $v =-1$ and $x=0$ to correspond to the horizon of black hole.
 Now we can expand around this point to get $v \simeq -1 +  \varkappa\, x$,
 where $\varkappa =  v'(0) = 1/4M$ is the surface gravity.
 Inserting this to equation (\ref{b}) we find that
\begin{equation}\label{c}
 \phi \sim \exp\left( i \frac\omega\varkappa\, \log(x) \right).
\end{equation}
To extract the positive and negative frequency parts as defined by
 freely falling observers near the horizon, we must analytically
continue the solution to negative $x$ through the upper and lower
complex $x$ plane. In this way we obtain two functions $\phi_+$
and $\phi_-$ corresponding to positive and negative wavevectors,
respectively:
\begin{equation}
\label{cc} \phi_+=\phi+\exp{(-\pi\frac{\omega}{\varkappa})}
\tilde\phi,\;\;\;\;
\phi_-=\phi+\exp{(\pi\frac{\omega}{\varkappa})}\tilde\phi,
\end{equation}
where $\tilde{\phi}$ is defined by
\begin{equation}
\label{ccc} \tilde{\phi}(x)=\left\{
\begin{array}{ll}
0&x>0\\
\phi(-x)&x<0.
\end{array}
\right.
\end{equation}
They agree with (\ref{c}) on the positive $x$ axis and their ratio
for $x<0$ equals
\begin{equation}\label{d}
  \frac{\phi_+}{\phi_-}= \exp\left( -2\pi\frac\omega\varkappa \right).
\end{equation}
To compute the average number of particles of energy $\hbar\omega$
produced in the Hawking process one has to evaluate the square of
the norm of the negative frequency part of the
 resulting initial mode.
The positive frequency final mode vanishing for the negative $x$
is equal
\begin{equation}
\phi\;\sim\psi=\phi_+ -
\exp{(-2\pi\frac{\omega}{\varkappa})}\phi_- .
\end{equation}
Using the fact that Klein-Gordon inner product $\Omega^{KG}$
satisfies
$$
\Omega^{KG}(\phi_+,\phi_+)=1-\exp{(-2\pi\frac{\omega}{\varkappa})},\quad
\Omega^{KG}(\phi_-,\phi_-)=1-\exp{(2\pi\frac{\omega}{\varkappa})},$$
\begin{equation} \Omega^{KG}(\phi_+,\phi_-)=0,
\end{equation}
one finds
$$
-<n_{\omega}>=\frac{\Omega^{KG}(\phi_-,\phi_-)\exp{(-4\pi\frac{\omega}{\varkappa})}}
{\Omega^{KG}(\psi,\psi)}=$$\begin{equation}\label{e}
=\frac{1}{\frac{\Omega^{KG}(\phi_+,\phi_+)}{\Omega^{KG}(\phi_-,\phi_-)}
\exp{(4\pi\frac{\omega}{\varkappa})}+1}=\frac{1}{-\exp{(2\pi\frac{\omega}{\varkappa})}+1}
\end{equation}
which is exactly the Hawking formula.

\section{Black hole radiation with $\kappa$-Poincare dispersion }

In the metric (\ref{4})
the field equation describing the dynamics of
 massless scalar field
is given by
\begin{equation} \label{fe}
\hat{S}\phi\equiv\left[\kappa^2(1-e^{-\frac
i{\kappa}(\partial_t+\partial_x\; v(x))}) (1-e^{-\frac i {\kappa}
(\partial_t+v(x)\;\partial_x )}) +
\partial_x^2\right]\phi = 0.\end{equation}
Because $\hat{S}$ is a self--adjoint differential operator in the
sense that
\begin{equation}
\int dt\, dx\, \psi_1^*\hat{S}\psi_2 = \int dt\, dx\,
(\hat{S}\psi_1)^*\psi_2
\end{equation}
for complex functions $\psi_1$, $\psi_2$ vanishing with all
derivatives at the boundary of the integration domain, it is justified to proceed with construction of
 the inner product as it was done in Section 2.

Now we again make use of the WKB approximation (\ref{b}) and
substitute it into eq.~(\ref{fe}) neglecting the terms $\partial_x
v$ and $\partial_x k/k$. As the result we obtain
\begin{equation}\label{7}
 k^2 =\kappa^2 (1-e^{-(\omega - v(x) k)/\kappa})^2 \equiv f(\omega-v(x) k) = f(\omega').
\end{equation}
Using this and equation (\ref{xf}), one sees that in
regions where $v$ is approximately constant, solutions satisfying
$i(\partial_t+v\partial_x)\phi=\omega'\phi$ have positive
(negative) Klein--Gordon norm for $\omega'>0$ ($\omega'<0$).

 The
intercept points of the line $\omega'=\omega - v(x) k$ with
$x$-dependent slope and the curve $k^2 - f^{2}(\omega')=0$
 on the $(k,\, \omega')$ plane correspond to possible values of wavevectors.
 In the case of $\kappa$-Poincar\'e dispersion relation (\ref{1a}), we have
\begin{equation}\label{2}
 \omega'   = -\kappa\log\left( 1 \pm
 \frac{|k|}\kappa\right) \equiv F(k)
\end{equation}
for super-luminal $\kappa>0$,  and for the sub-luminal,  $\kappa
<0$ cases (see Figures 1 and 2 for details.)
\newline

 Let us pause at this point to make an important comment.
 Any reasonable function $F(k)$ must satisfy the condition $$F(k) \sim k$$ for sufficiently
 small $k$, so that it corresponds to the standard dispersion relation for small momenta. To make this statement more precise, it should be noted that from purely dimensional
  reason, the function $F(k)$ must contain a scale, so it is of the form $F(k;\kappa)$ and
  satisfies the condition
  $$F(k;\kappa) = k + O\left(\frac{k^2}{\kappa^2}\right) \quad \mbox{for $k/\kappa \ll 1$}.$$
   This means that for $\omega$ small enough (i.e., $\omega/\kappa \ll 1$) the solutions of
    eq.~(\ref{7}) is in leading order the same as in the standard case. One can conclude
    therefore that, if one defines the initial vacuum as seen by the freely falling observer
     near the horizon, for low frequencies the spectrum will be almost thermal with  differences in
    the high frequency part of the spectrum. If the qualitative thermal behavior  holds for
     high frequencies  as well, then the modifications due to the deformed dispersion
     relation would be  exponentially suppressed, at least for temperatures
     (surface gravity) small compared to the scale $\kappa$. But, of course, it
      does not make  sense at all to consider situation when the temperature
      is of order of, or higher than $\kappa$, because such a regime corresponds
      to quantum gravity (recall that $\kappa$ is of order of Planck scale) and the
      approximation of Schwarzschild background geometry would almost certainly not hold.

The question then arises if for temperatures reasonably below $\kappa$ scale there is
any deviation from the Hawking result\footnote{At this point it is
 worth recalling that the similar question has been asked in the context
 of inflationary cosmology (i.e., is it any deviation from Harrison--Zeldovich
 spectrum if one makes use of modified dispersion relation  \cite{mb}--\cite{niemp}),
 and it turned out that the answer depends on the form of relation used, as well as the
 form of initial conditions.}. It should be stressed that the rationale for asking this question is rather different
   from that in \cite{unruh} -- \cite{bhlaser}. There the question was if, by making use
   of a non-standard dispersion relation, one can avoid trans-Planckian frequencies keeping
   at the same time the qualitative picture of Hawking process. Here our goal   is different:
    we have the dispersion relation to start with, and the question we ask ourselves is if it
    does change the thermal behavior of black holes?
\newline

 To answer this question consider  the group velocity of a wavepacket. It can be expressed as
\begin{equation}\label{8}
  \upsilon_g = \upsilon'_g + v(x),
\end{equation}
where $\upsilon_g = d\omega/dk$ is the group velocity with respect
to the static frame  and $\upsilon'_g = d\omega'/dk$ is the one
corresponding to the freely falling frame. The detailed discussion
presented in \cite{corjac} indicates
 that the {\em sin equa non} condition for Hawking radiation to occur is that there exists an
 ingoing mode (i.e., a mode with $\upsilon_g<0$) with {\em negative} free-fall frequency
 $\omega'$ for positive Killing frequency $\omega$ outside the
 horizon
 (i.e., for $v(x) > -1$).

Knowing this we can turn to the analysis of the
$\kappa$-Poincar\'e dispersion relation.
 Let us consider first the super-luminal case ($\kappa>0$).
 The corresponding picture can be found in Figure 1.

\begin{figure}[ht]
\begin{center}
\includegraphics[width=7cm]{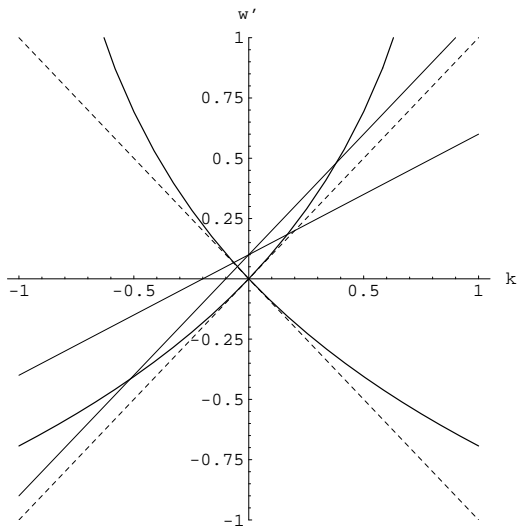}
\end{center}
\caption{The behavior $\omega'$ vs.~$k$ in  super-luminal case
($\kappa = 1$).}
\end{figure}

We see that  there are three  intersections of the line $\omega - v(x) k$ with the
curve $F(k)$, eq.~(\ref{2}) which, following notation of
\cite{corjac}, we call (from the left to the right of the figure)
$k_-$, $k_{-s}$, $k_+$. For these intersections one can estimate
the signs of group velocities in both frames of reference
\begin{equation}
\begin{array}{l l l l}
\mathrm{for}\; k_- & v_g' >0& v_g<0 & \rightarrow \; \mathrm{ingoing \; packet} \\
\mathrm{for}\; k_{-s} & v_g' <0& v_g<0 & \rightarrow \; \mathrm{ingoing \; packet} \\
\mathrm{for}\; k_+ & v_g' >0& v_g>0 & \rightarrow \;
\mathrm{outgoing \; packet}.
\end{array}
\end{equation}
  Following the standard analysis (see also \cite{corjac}) we
conclude that the number of created particles in final state
$\psi_{out}$ is given by
\begin{equation}
n(\psi_{out})=-\Omega(\psi_-,\psi_-),
\end{equation} where $\psi_-$
denotes the ingoing part of the solution with negative free-fall
frequency. In our case the solution corresponding to the  wavevector $k_-$
satisfies this condition. However we do not know whether it is
possible for the outgoing solution centered around $k_+$ to
undergo the mode conversion which is essential in the analysis.
Although  the naive diagram analysis indicates that it is not the
case, one must remember that near the horizon the WKB approximation
brakes down and the mode conversion as well as the particle production might be possible.
 Here, unlike in \cite{cor} and
\cite{bhlaser}, the negative  frequency modes that give rise to
particle production originate at infinity rather than inside
the horizon.
Still, in order to reach any final conclusion, one needs to perform
detailed numerical studies.
\newline

\begin{figure}[ht]
\begin{center}
\includegraphics[width=7cm]{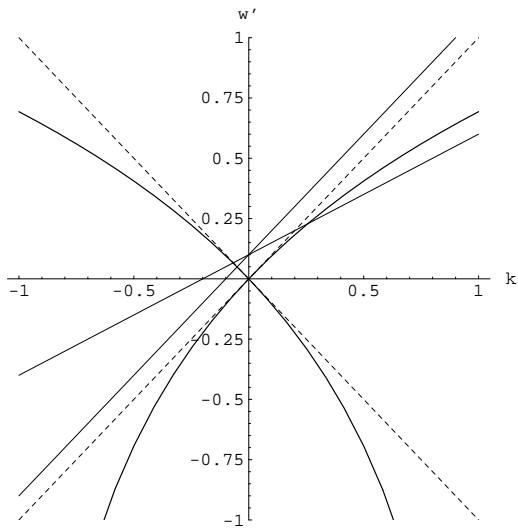}
\end{center}
\caption{The behavior $\omega'$ vs.~$k$ in  sub-luminal case
($\kappa=-1$).}
\end{figure}

Let us now turn to the sub-luminal case, illustrated in Figure 2.
In this case for  given positive $\omega$ outside the horizon
there are, depending on the distance from the horizon, either
three solutions   $k_{-s},\; k_{+s},\;k_+$ (ingoing, outgoing,
ingoing mode respectively), or two  $k_{-s},\;k_{+s}$
(ingoing, hanging), or only one  $k_{-s}$. None of them
contributes to the creation of particles since all have a positive
free-fall frequency. Again this conclusion is drawn under the
assumption of the validity of WKB approximation which for small
Killing frequencies $\omega$ brakes in the vicinity of the horizon.
Taking sufficiently small $\omega$  one can get arbitrarily close
to the horizon. Then the mode conversion to negative $\omega'$
might be possible resulting in two negative wavevector
wavepackets, one of which would represent the ingoing mode. Such negative frequency wavepacket  giving rise to the black hole
radiation would originate far inside the horizon. Note, however,
that there is a difference between the subluminal dispersion case
considered by \cite{corjac} and the one discussed here. The
intersection with the negative branch of dispersion relation
occurs only inside the horizon. The appearance of a "gap" makes
the mode conversion less obvious.
\newline

In the analysis above we followed exactly the methods developed by
Corley and Jacobson \cite{corjac}, but this is only half of a
story. In our case dispersion relations are not symmetric with
respect to inversion $\omega'\rightarrow \, -\omega'$, and
therefore we have to do with complex field equations. Since the
corresponding field operator $\hat{\phi}$ is not hermitian
one might, in analogy with the standard analysis, define creation
and annihilation operators for particles and antiparticles, the
latter corresponding to negative frequency waves.

Therefore the complete analysis of the spontaneous creation of
particles (see for example \cite{gibbons}, \cite{wald}) should
include tracing both the outgoing positive frequency  and outgoing
negative frequency modes backwards in time, and will be presented
elsewhere \cite{arek}.
\newline

 We have shown therefore
that in both cases the existence of "Hawking-type" radiation is
possible.
However, in both super- and subluminal regimes there are
obstacles that, in principle, may lead to strong suppression of
the radiation for higher frequencies. Of course both these results
rely on the validity of WKB regime, and the detailed analysis of
this regime as well as the quantitative numerical study of the
qualitative results of this paper will be presented elsewhere
\cite{dobno}. The analysis presented here indicates however that
such studies are worth undertaking, since the qualitative
picture which  emerged from our investigations presented here
differs in many respects from the one studied before.
\newline

{\bf Acknowledgement}. We would like to thank Ted Jacobson for discussion and comments.

\end{document}